\documentclass[useAMS,usenatbib]{mn2e}
\newcommand{\beq}{\begin{equation}}
\newcommand{\eeq}{\end{equation}}

\newcommand{\apj}{ApJ}
\newcommand{\apjl}{ApJL}
\newcommand{\apjs}{ApJS}
\newcommand{\aj}{AJ}
\newcommand{\mnras}{MNRAS}
\newcommand{\aap}{A\&A}

\newcommand{\araa}{ARA\&A}

\newdimen\hssize
\newdimen\hdsize
\usepackage{amssymb} % for \grtsim

\newcommand{\Hi}{\rm H~{I}}
\newcommand{\Hii}{\rm H~{II}}

\newcommand{\Oi}{[\rm O~ \sc{I}]}
\newcommand{\Oii}{[\rm O~ \sc{II}]}
\newcommand{\Oiii}{[\rm O~ {III}]}
\newcommand{\Nii}{[\rm N~ \sc{II}]}
\newcommand{\Sii}{[\rm S~ \sc{II}]}

\usepackage{graphicx}
\usepackage{color}
\usepackage{ulem} % for \grtsim

\title[metallicity estimation of SF ETGs]{Estimating the 
metallicity of star-forming early-type galaxies}
\author[Wu \& Zhang]{Yu-Zhong Wu \thanks{E-mail:yzwu@nao.cas.cn} \& Wei Zhang \\
   ${}$CAS Key Laboratory of Optical Astronomy, National
Astronomical Observatories, Beijing, 100101, China\\}

\begin{document}

\date{Accepted ........ Received ........; in original form ........}
\pagerange{\pageref{firstpage}--\pageref{lastpage}} \pubyear{2020}
\maketitle
\label{firstpage}

\begin{abstract}

We derive data of 4615 star-forming 
early-type galaxies (ETGs), which come from 
cross-match of the $Galaxy~Zoo~1$ and the catalogue of the 
MPA-JHU emission-line measurements for the Sloan Digital Sky 
Survey Data Release 7. Our sample distributes mainly at 
$\rm -0.7<log(SFR[M_{\sun}yr^{-1}])<1.2$, and the median value 
of our SFRs is slightly higher than that shown in 
Davis \& Young. We display a significant trend of lower/higher 
stellar mass ETGs to have lower/higher SFR, and obtain our 
sample best fit of 
log(SFR)=$ (0.74\pm0.01)$log$(M_{*}/M_{\sun})-(7.64\pm0.10)$, 
finding the same slope as that found in Cano-D\'{i}az et al. 
In our star-forming ETG sample, we demonstrate 
clearly the correlation of the stellar mass and metallicity (MZ) 
relation. We find that higher 
metallicity measurements may be introduced by the diffuse 
ionized gas, when the D16, Sanch18, and Sander18 indicators 
are used to calibrate the metallicity of 
ETGs. We show the relations between SFR and 12+log(O/H) 
with different metallicity estimators, and suggest that their 
correlations may be a consequence of the SFR-stellar mass 
and MZ relations in ETGs.

\end{abstract}


\section{INTRODUCTION}

Some galaxy surveys reveal the two different locales in the 
colour-stellar magnitude diagram, the `red cloud' and the 
`blue cloud' (Baldry et al. 2004). The blue cloud contains 
blue, star-forming late-type galaxies (disc-dominated), while 
the red cloud includes red, passive early-type galaxies 
(ETGs; elliptical, S0, and bulge-dominated spiral). In optical 
colour-magnitude/mass diagram, the red cloud generally 
denotes ETGs. Based on a visual classification of images of 
Sloan Digital Sky Survey (SDSS) 
galaxies, ETGs contain morphologically elliptical and lenticular 
galaxies, and it is suggested that galaxies are `red and dead' 
objects.

Ultraviolet (UV) observations confirm recent star formation 
(SF) in many ETGs, and the $Galaxy ~Evolution~ Explorer$ (GALEX) 
and $Hubble~Space~Telescope~$(HST) provide a good means to 
trace detection of SF in ETGs. Using the GALEX photometric 
data, Yi et al. (2005) have showed that the near-UV colour-magnitude 
relation is a perfect instrument, tracing the late SF history, 
and that it demonstrates signs of recent SF in ETGs. 
Combining the SDSS data release 3 (DR3) and GALEX Medium 
Imaging Survey, Kaviraj et al. (2007) found some recent SF 
in $\ge 30\%$ of the ETG sample of 
2,100 galaxies during $\lesssim 1$ Gyr. Studying NGC 4150 
near-UV/optical data from the Wide Field Camera 3 on the HST, 
Crockett et al. (2011) exhibited recent SF, fueled 
by a large reservoir of molecular gas induced by minor merger,
and suggested that $\sim 2-3\%$ of the stellar mass is 
contributed by the SF.

The gas that fuels low-level SF has various sources, 
and it can be classified into internal and 
external processes. The internal mechanism contains stellar 
mass-loss, while the external process includes mainly accretion of 
cold gas, hot gas and clumpy accretion (Bryant et al. 2019). 
Using the CO detections of ETGs in single-dish surveys, 
Young (2005) suggested that stellar mass loss could have 
resulted in the molecular gas in some ETGs, and also found 
some evidence that the molecular gas is traced from both 
internal and external processes in NGC 83 and NGC 2320. Using 
the GALFORM model of galaxy formation, Lagos et al. (2014) 
found that radiative cooling from the hot halos of ETGs 
provides neural gas contents in a majority of ETGs at 
$z\approx0$. Obtaining \Hi~ synthesis imaging data of three 
ETGs, Chung et al. (2012) investigated their gas environment
and  \Hi~ properties, and found some \Hi~gas directly from 
neighbours, which is evidence for cold gas accretion.
Considering gas inflows in various mergers or interactions 
between galaxies, clumpy accretion is also an external
process.

Late-type galaxies can quench SF if they lack
enough supply of various gas, and their sites will change 
from the blue cloud to the red cloud on the 
colour-magnitude diagram (e.g. Faber et al. 2007; 
Cortese \& Hughes 2009). At the same time, much evidence 
shows that the stellar 
mass in red cloud galaxies has 
been increasing (e.g. Bell et al. 2004; Brown et al. 2007). 
As a result of various major mergers, minor mergers, and the 
inflow of the intergalactic medium, ETGs can accrete various gas, 
triggering SF, which will then drive the ETGs to migrate 
from the red cloud to the blue cloud on the 
colour-magnitude diagram.

\begin{figure}
\begin{center}
\includegraphics[width=8cm,height=6cm]{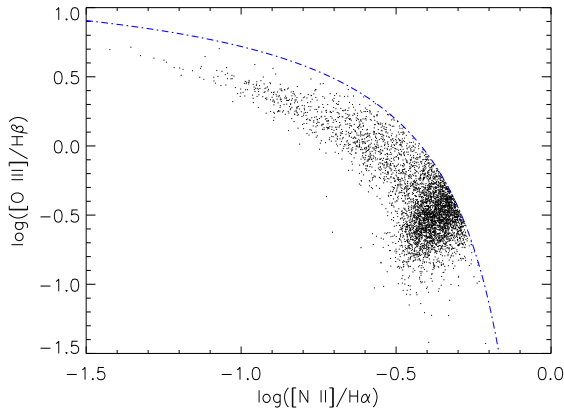}
\caption{Traditional diagnostic diagram. The black dots are
star-forming ETGs. The blue dotted-dashed line shows the 
semi-empirical lower limit of Kauffmann et al. (2003a) for 
SFGs.} 
\end{center}
\end{figure}

Schawinski et al. (2009) used the SDSS DR6 and u-r colour data 
to obtain 204 star-forming blue ETGs with the help of the visual 
morphology from the Galaxy Zoo. These ETGs have significantly 
more blue u-r colour than the red cloud galaxies with higher 
SFRs ($0.5<$ SFR$[M_{\sun}~\rm yr^{-1}]<~50~$), and their sample 
provides SFRs and emission line classification. Utilizing the 
SDSS DR7 and Galaxy Zoo 1, Davis \& Young (2019) derived a 
larger sample of star-forming 
elliptical galaxies - based on 
the standard of Kauffmann et al. (2003a), on the 
Baldwin-PHilips-Terevich diagram (BPT, Baldwin et al. 1981) 
- with the metallicity measurements. The median value is 
$2~M_{\sun}~\rm yr^{-1}$ in their ETG sample.

Several works claim to explore the properties of ETGs, and 
they demonstrate that these ETGs have significant SFR measurements.
In this study, we mainly focused on the SDSS DR7 and Galaxy Zoo 
1 to derive star-forming ETG sample. We also study the 
distributions of various parameters and compare them with 
star-forming galaxies (SFGs) or composite ETGs. (These 
galaxies are defined as the ETG sample, which lies in 
the composite region on the BPT diagram, and these ETGs are 
called composite ETGs in this paper). In addition, 
we estimate their metallicities with six abundance indicators. 
In Section 2, we describe the ETG sample and data in detail. 
We present sample properties of our star-forming ETGs in 
Section 3. In Section 4, we calibrate their metallicities 
with six abundance estimators, and describe metallicity 
properties of our ETG sample. A summary of our results is 
shown in Section 5.

\section{THE DATA}

In this study, our sample is chosen from the data of the SDSS 
Seventh Data Release (DR7, Abazajian et al. 2009). In the SDSS 
DR7 catalogue of Max Planck Institute for Astrophysics $-$ John 
Hopkins University (MPA-JHU), we can utilize measurements of 
emission line fluxes, stellar masses, redshifts, and SFRs 
of about 900,000 galaxies, and they are publicly available.

\begin{figure}
\begin{center}
\includegraphics[width=8cm,height=6cm]{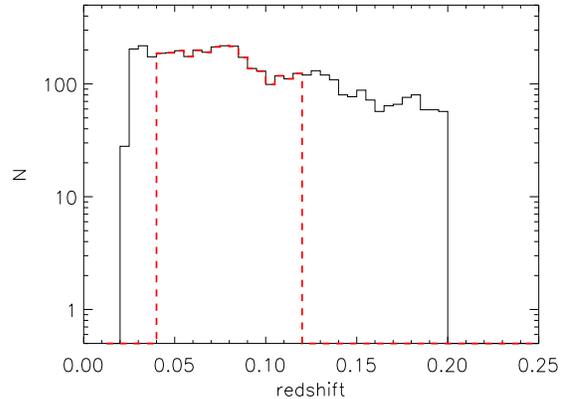}
\caption{Distributions of redshift for star-forming ETGs. The 
black solid and red dashed lines display the samples of 
star-forming ETGs with $0.023<z<0.2$ and $0.04<z<0.12$, 
respectively.}
%%{\noindent  \vglue 0.5cm {\sc  }}
\end{center}
\end{figure}

\begin{figure*}
\begin{center}
\includegraphics[width=8cm,height=6cm]{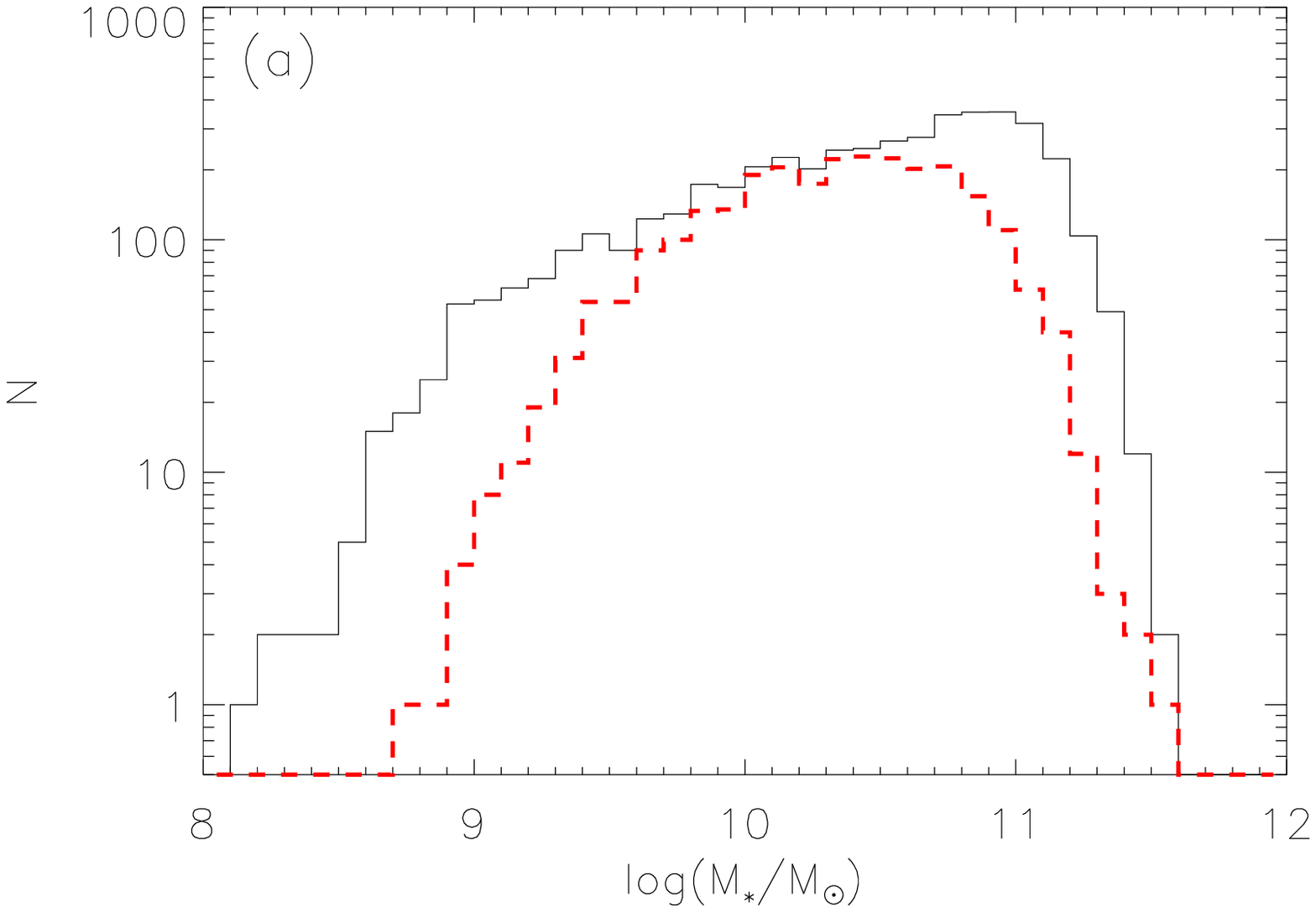}
\includegraphics[width=8cm,height=6cm]{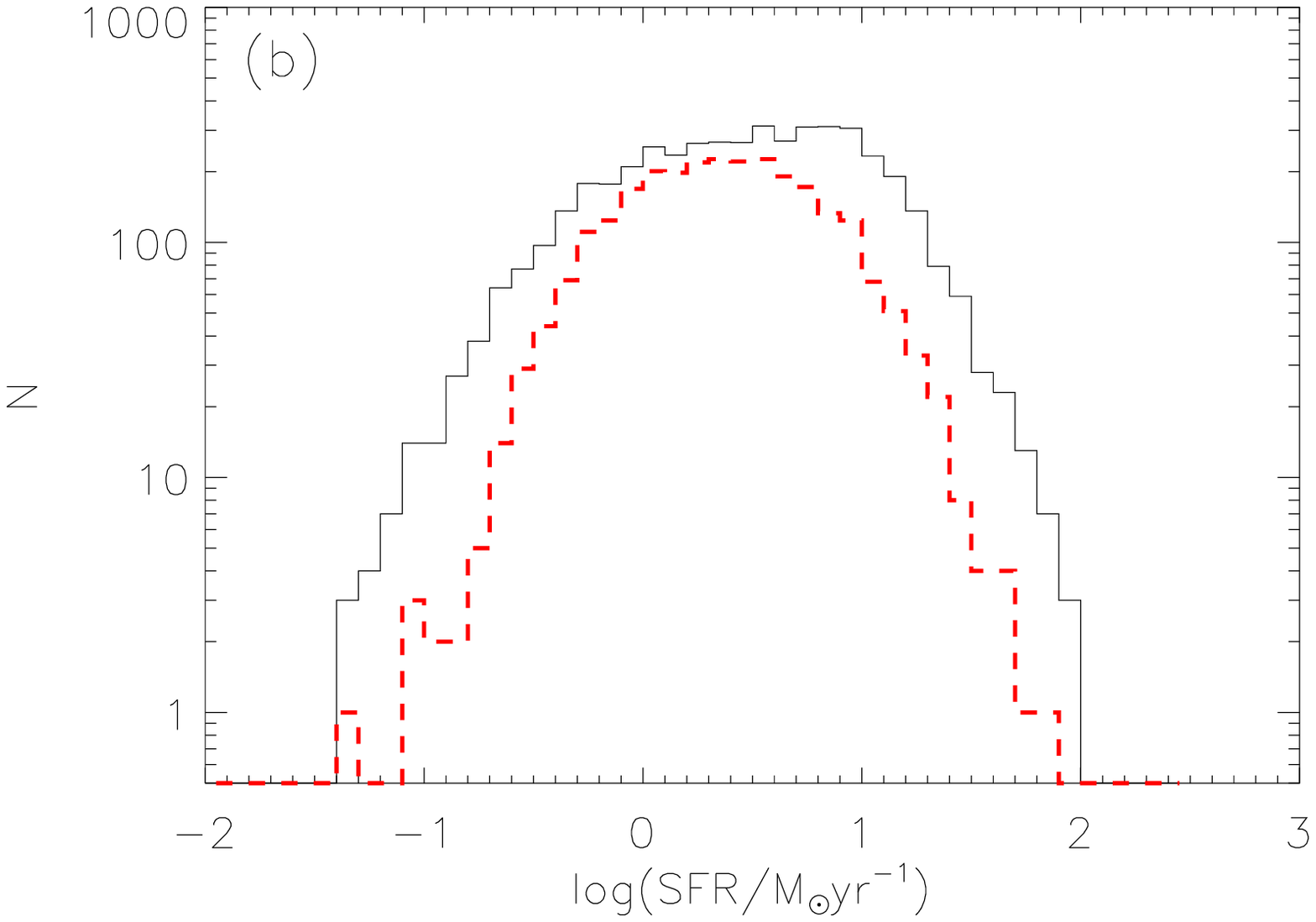}
\caption{Comparison of stellar mass (left-hand panel) and SFR 
(right-hand pane) distributions for star-forming ETGs with 
different redshift ranges. The black and red dashed lines are 
the same as Figure 2.}
%{\noindent  \vglue 0.5cm {\sc  }}
\end{center}
\end{figure*}

In this work, we study the star-forming ETGs, and we 
adopt a similar method to that of Wu (2020) to select
our sample, before the BPT diagram is used 
to choose the composite ETG sample.
Based on a wavelength range of 3800-9200 {\AA} of the SDSS 
spectra, we select the lower limit ($z\approx 0.023$) of 
redshifts in our sample, and we can be assured that 
$\Oii~\lambda \lambda$ 3227, 3229 will appear in the observed range 
(Wu \& Zhang 2013). The upper limit of redshifts is $0.2$ in our 
study. We need to restrict the aperture-covering fraction of $>20\%$, 
and the fraction is computed from the fiber and Petrosian magnitudes 
in the r band. We select these galaxies, which have a signal-to-noise ratio 
(S/N) $>3$ for H$\alpha$ and S/N $>2$ for $\Oii~\lambda \lambda$ 3227, 
3229, and $\Nii~\lambda 6584$. Regarding the SFR measurement 
in the catalogue, we require an SFR FLAG keyword, describing the 
status of the SFR measurements, to be equal to zero. In 
consequence, 231,666 galaxies constitute our initial sample.

For the ETG selection, we take the two judgment standards to
select our ETGs (Wu 2020). One criteria is that the galaxies 
should satisfy $n_{\rm Sersic}>2.5$, and the other one is that 
the elliptical probability is larger than 0.5 
(Herpich et al. 2018), which comes from Galaxy Zoo 
1 (Lintott et al. 2008, 2011). 
The measurement of the S\'{e}rsic index 
is chosen from the New York University Value-Added Galaxy 
Catalogue (NYU-VAGC, Blanton et al. 2005). 
First, we cross-match the NYU-VAGC with our initial sample
within $2''$, and we choose galaxies with $n_{\rm Sersic}>2.5$.
Then we use the elliptical probability of $>0.5$ to select
ETGs (Herpich et al. 2018), and match these 
$n_{\rm Sersic}>2.5$ galaxies with those found in table 2 
of Lintott et al. (2011) within
$2''$. Finally, we derive an ETG sample of 28,093 objects.

In this paper, we study these ETGs, which come from the 
above-mentioned ETG sample and also lie in the $\Hii$~ region 
on the BPT diagram: they are called as star-forming ETGs. 
The topic of ionised gas has been 
discussed in many recent articles, and in particular in 
the two recent reviews on the topic by 
S\'{a}nchez 2020 (Figures 1 and 5, and main text) and 
S\'{a}nchez et al. 2020 (Figures. 2-6, section 4).
To obtain a reliable ETG sample, we cannot define definitive 
boundaries only using the BPT diagram, and other additional 
parameters are required. The ionisation can be studied and 
discussed through the global and local dominance 
(Stasi\'{n}ska et al. 2008; Cid Fernandes et al. 2010; 
S\'{a}nchez et al. 2014; Lacerda et al. 2018, 2020). In 
general, if the EW(H$\alpha$)$<6$\AA~(for the SDSS aperture), 
it is difficult to justify that the observed ionisation is due 
to SF (Stasi\'{n}ska et al. 2008; Cid Fernandes et al. 2010).
Also, the velocity dispersion may be a better classification of 
the ionisation (not for the SDSS single fiber, D'Agostino et al. 
2019). Finally, other line ratios should be considered, as 
the use of the BPT diagram in addtion to the EW(H$\alpha$) is 
not enough to exclude other ionising sources, such as low-metallicity 
weak AGNs and shocks. Shock ionisation is nowadays known to be 
present in ETGs (Dopita et al. 2016; Cheung et al. 2016; 
L\'{o}pez-Cob\'{a} et al. 2020). Its EW(H$\alpha$) can take any 
value, and can be observed in any location in the BPT diagram. 
Moreover, the use of other line ratios, such as $\Sii/\rm H\alpha$ and 
$\Oi/\rm H\alpha$, can improve the classification of the ionisation 
(Kewley et al. 2001; S\'{a}nchez 2020). We select our sample 
ionised by SF by adopting three 
conditions: 1, the Kauffmann et al. (2003a) line for the BPT
diagram; 2, the Kewley et al. (2001) demarcation lines in the
$\Sii/\rm H\alpha$ vs $\Oiii/\rm H\beta$ and
$\Oi/\rm H\alpha$ vs $\Oiii/\rm H\beta$; 3, the H$\alpha$ 
equivalent width (EW(H$\alpha$)) 
of $>6$\AA~ \textbf(Cid Fernandes et al. 2010). 
As these galaxies are dominated by SF activity, we 
consider contribution of SF on photoionization from 
these galaxies. Here, we obtain 4,615 star-forming ETGs.

In the measurements of $M_{*}$ (Kauffmann et al. 2003b)
and SFR (Brinchmann et al. 2004), we adopt 
the total stellar mass and total SFR from the MPA-JHU catalogue.
The mass-to-light ratios and the z-band attenuation values 
within the SDSS fiber are extrapolated to the whole galaxy, 
under the assumption that the fiber emission line contribution 
is the same as the global one. The total mass is calculated by 
multiplying the dust-corrected and K-corrected luminosity of 
the galaxy by the mass-to-light ratio estimate. The SFR is 
derived using the H$\alpha$ flux within the SDSS fibers, and 
fiber correct using the galaxy colours is purely empirical.
All aperture bias in SFR estimates can be corrected in 
statistical way (Brinchmann et al. 2004). In addition, we need to 
correct a Chabrier (2003) initial mass function (IMF) from a 
Kroupa (2001) IMF assumed in the MPA-JHU catalogue for the SDSS 
DR7. Because our ETG sample is selected by using the above-mentioned 
three conditions, we can use various metallicity calibrators to estimate 
the gas-phase oxygen abundance. The O3N2 index introduced by Alloin et al. 
(1979) depends on $\Oiii$/H$\beta$ and $\Nii$/H$\alpha$. One 
of the most popular calibrations relate the index and the 
metallicity using $12+\rm log(O/H)=8.73-0.32 \times O3N2$ proposed by 
Pettini \& Pagel (2004). There have been several updates of this 
calibrator (Nagao et al. 2006; P\'{e}rez-Montero \& Contini 2009). 
The calibrator has been recently updated by 
Marino et al. (2013), and we adopt the calibrator in this study.
Also, we use the following various metallicity calibrators: D16, the 
N2S2H$\alpha$ method of Dopita et al. (2016); Jon15, the O32 
method of Jones et al. (2015); Curti17, the O3S2 method of Curti 
et al. (2017); Sanch18, the N2 method of 
S\'{a}nchez-Almeida et al. (2018); Sander18, the N2O2 method 
of Sanders et al. (2018).

\section{Sample properties of star-forming early-type galaxies}

\begin{table*}
\caption{Sample of the star-forming ETGs.}
%\begin{scriptsize}
\begin{tiny}%{small}
\begin{center}
\setlength{\tabcolsep}{4.5pt}
\renewcommand{\arraystretch}{1.2}
\begin{tabular}{cccccccccccccl}\hline \hline  
R.A. & Decl. & Redshift & {log($M_{*}$)} &
 {log(SFR)}  & $\rm frac^a$ & \multicolumn{6}{c} {$12+\rm log(O/H)$}  &  \\
\cline{7-12}

(J2000)&(J2000)& & $M_{\sun}$ & $M_{\sun}/yr$ &  & Ma13 & Jon15 & Curti17& D16 & Sanch18 &Sander18 \\

(1)& (2) & (3) & (4) &(5)&(6) &(7)&(8) &(9)&(10)&(11)&(12) \\
\hline

   11 39 37.2 &  21 18 05.4 & 0.16 & 10.95 & 1.43 & 0.38 &  $8.59\pm0.02$ & $8.58\pm0.04$ & $8.81\pm0.02$ & $9.02\pm0.03$ & $8.80\pm0.01$ & $9.31\pm0.03$\\
   13 43 02.6 &  18 52 21.0 & 0.08 & 10.46 & 0.41 & 0.44 &  $8.58\pm0.01$ & $8.64\pm0.03$ & $8.76\pm0.01$ & $8.75\pm0.02$ & $8.78\pm0.01$ & $9.12\pm0.02$\\
   13 54 58.1 &  60 30 54.0 & 0.09 & 10.40 & -0.07& 0.50 &  $8.57\pm0.02$ & $8.63\pm0.05$ & $8.75\pm0.02$ & $8.83\pm0.04$ & $8.80\pm0.02$ & $8.96\pm0.04$\\
   08 18 32.2 &  29 48 03.6 & 0.08 & 10.58 & 0.39 & 0.32 &  $8.56\pm0.02$ & $8.61\pm0.05$ & $8.79\pm0.02$ & $8.90\pm0.05$ & $8.74\pm0.01$ & $9.10\pm0.04$\\
   15 35 24.7 &  08 58 08.4 & 0.05 &  9.77 & -0.15& 0.42 &  $8.54\pm0.01$ & $8.61\pm0.02$ & $8.74\pm0.01$ & $8.71\pm0.01$ & $8.74\pm0.01$ & $8.99\pm0.02$\\
   08 01 38.4 &  19 34 49.8 & 0.12 & 10.91 & 0.87 & 0.31 &  $8.59\pm0.02$ & $8.63\pm0.06$ & $8.83\pm0.02$ & $8.93\pm0.04$ & $8.74\pm0.01$ & $9.17\pm0.03$\\
   13 58 18.7 &  06 25 53.3 & 0.18 & 10.85 & 1.51 & 0.39 &  $8.47\pm0.01$ & $8.35\pm0.03$ & $8.67\pm0.01$ & $9.01\pm0.04$ & $8.75\pm0.01$ & $9.25\pm0.03$\\
   15 08 04.1 &  04 04 49.2 & 0.09 & 10.82 & 1.27 & 0.44 &  $8.53\pm0.00$ & $8.56\pm0.01$ & $8.73\pm0.00$ & $8.87\pm0.01$ & $8.79\pm0.00$ & $9.06\pm0.01$\\
   15 05 01.7 &  08 20 12.4 & 0.09 & 10.16 &-0.48 & 0.50 &  $8.41\pm0.01$ & $8.53\pm0.02$ & $8.55\pm0.02$ & $8.44\pm0.05$ & $8.69\pm0.02$ & $8.56\pm0.03$\\
   13 19 02.6 &  08 14 29.6 & 0.13 & 10.95 & 0.04 & 0.27 &  $8.48\pm0.03$ & $8.59\pm0.05$ & $8.64\pm0.03$ & $8.63\pm0.05$ & $8.77\pm0.03$ & $8.89\pm0.05$\\

...  & ...& ...  & ...  & ...   & ...& ...   & ...   &... & ...  & ... & ...  &         \\

\hline \hline

\end{tabular}
%\parbox{6.5in}
\parbox{7.0in}
{\baselineskip 11pt \noindent \vglue 0.5cm {\sc Note}: 
$^a$ Values were calculated from the fiber flux and petro flux ratio.}
\end{center}
\end{tiny}%{small}
%\end{scriptsize}
\end{table*}

In this section, we describe the various properties of the star-forming 
ETG sample. In Figure 1, we show these star-forming ETGs on the BPT 
diagram. The Kauffmann et al. (2003a) semi-empirical lower boundary 
for SFGs is shown by the blue dotted-dashed curve. Our 
sample of star-forming ETGs is displayed by the black dots. We can 
see that most of our star-forming ETGs are located in a small region on 
the BPT diagram.

In Figure 2, the redshift distributions of our star-forming ETG 
sample are shown. The red dashed and black lines represent the 
distributions of ETGs with $0.04<z<0.12$ and $0.023<z<0.2$, 
respectively. The former sample size accounts for $\sim 58\%$ of 
the latter one. We can see that the sample size does not changes 
clearly from redshift $z=0.025$ to $z=0.07$, and that the size 
generally decreases with increasing redshift at $z\gtrsim 0.08$. 
This shows that the size of our ETG sample is significantly 
affected by redshift, and indicates that the 
star-forming ETG sample with $ 0.023<z<0.2$ can be used to study 
various properties of ETGs.

Figure 3(a) describes the distributions of stellar mass for
the star-forming ETG samples. The black solid and red dashed lines 
represent the distributions of star-forming ETGs with 
$0.023<z<0.2$ and $0.04<z<0.12$, respectively. Compared with the 
distribution (occupying mainly the lower stellar mass part of the 
ETGs with $0.023<z<0.2$) of stellar mass for composite ETGs 
(Wu 2021), the $0.04<z<0.12$ star-forming 
ETGs take up the centre region of the distribution of stellar 
mass for star-forming ETGs with $0.023<z<0.2$. In the 
$0.04<z<0.12$ star-forming ETG sample, it distributes mainly at 
$9.5 <$log$(M_{*}/M_{\sun})<11.0$, accounting for $91\%$ of the 
ETG sample. The star-forming ETGs with $0.023<z<0.2$ distribute 
mainly $9.0 <$log$(M_{*}/M_{\sun})<11.5$, occupying $97\%$ of
our ETG sample. This indicates 
that star-forming ETGs will have 
different main distribution range of stellar mass when we select 
the ETG sample with different redshift ranges. We find that both 
star-forming ETG samples have a broader range of distribution for 
stellar mass. The ETG sample with $0.04<z<0.12$ has a median 
value of log$(M_{*}/M_{\sun})=10.36$, and the star-forming ETGs 
with $0.023<z<0.2$ present a median value of 
log$(M_{*}/M_{\sun})=10.50$. Compared with the median value, 
log$(M_{*}/M_{\sun})=10.75$, of stellar masses in composite 
ETGs (Wu 2021), the median value of star-forming ETGs 
decrease by $\sim 0.25$ dex.

\begin{figure}
\begin{center}
\includegraphics[width=8cm,height=6cm]{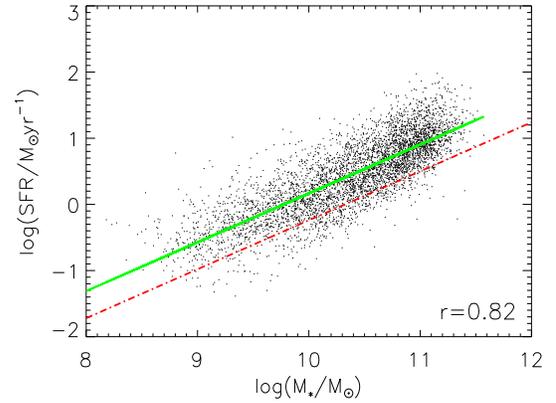}
\caption{ The SFR-M relation for star-forming ETGs. The red 
dotted-dashed and green lines represent the best 
fits of Cano-D\'{i}az et al. (2019) and this study for their 
corresponding data, respectively.}
\end{center}
\end{figure}

In Figure 3(b), we show the distribution of SFR for star-forming 
ETG samples. The distributions of SFR for $0.023<z<0.2$ and 
$0.04<z<0.12$ star-forming ETGs are displayed by the black solid 
and red dashed lines, respectively. The distribution of 
star-forming ETGs with $0.04<z<0.12$ occupies the lower SFR 
section of the ETGs with $0.023<z<0.2$. In the $0.04<z<0.12$ 
star-forming ETG sample, $\rm -0.5<log(SFR[M_{\sun}~yr^{-1}])<1.0$ 
is the main distribution of SFR, accounting for $91\%$ of the ETG 
sample. The $0.023<z<0.2$ star-forming ETG sample distributes 
mainly at $\rm -0.7<log(SFR[M_{\sun}~yr^{-1}])<1.2$, taking up 
$91\%$ of our ETG sample. 
The median values of the former and latter ETG samples are 
$\rm 2.3~M_{\sun}~yr^{-1}$ and $ 3.1~M_{\sun}~\rm yr^{-1}$, 
respectively. Compared with the median value 
($2~M_{\sun}~\rm yr^{-1}$) of SFRs of Davis \& Young (2019), our 
star-forming ETG sample has a slightly higher median value of SFRs 
than that shown in Davis \& Young (2019).

\begin{figure*}
\begin{center}
\includegraphics[width=16cm,height=12cm]{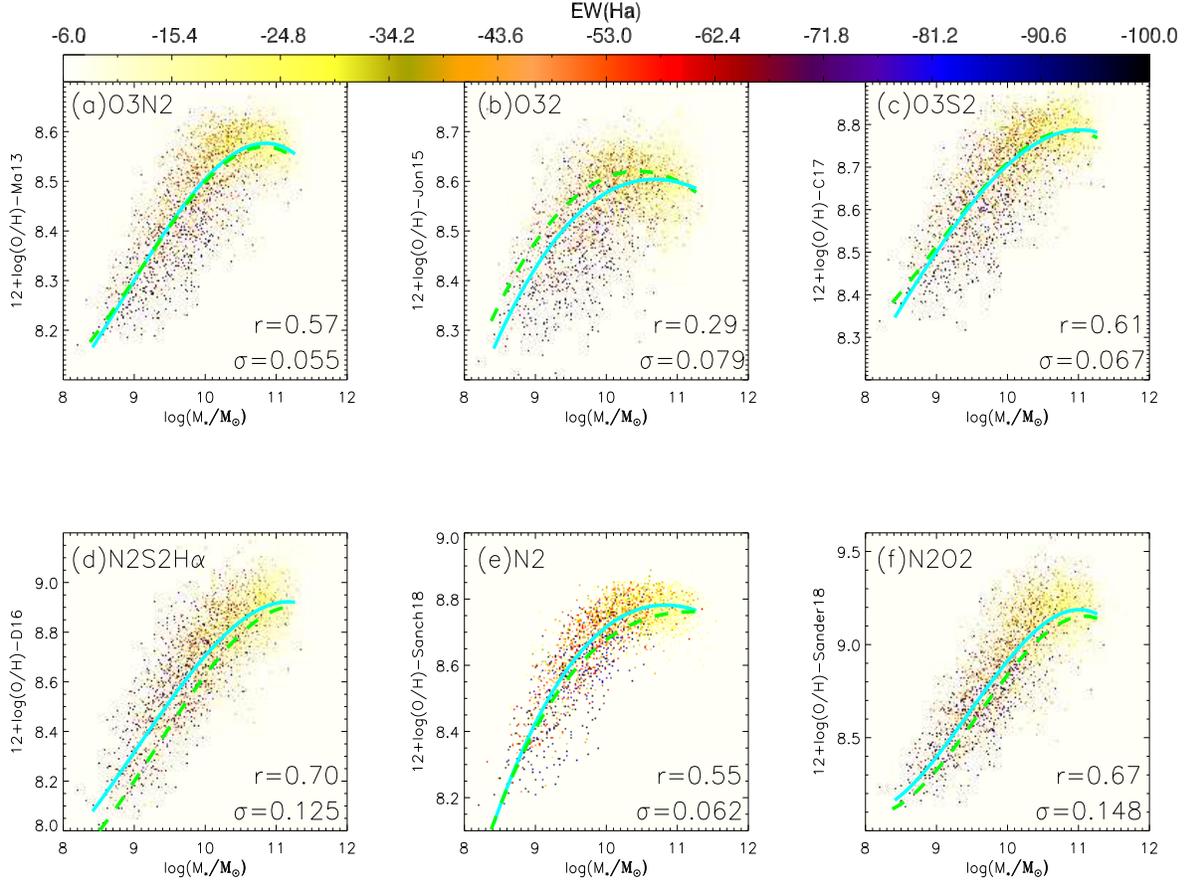}
\caption{Comparison of the MZ relations of star-forming
ETGs using the colour bar of EW(H$\alpha$), 
and their abundance estimators are the Ma13, Jon15, Culti17, 
D16, Sanch18, and Sander18, respectively. 
The green dashed and cyan solid lines
show the polynomial fits of data from 148,824 SFGs (see 
Section 2) and star-forming ETGs, 
respectively.}
\end{center}
\end{figure*}

In Figure 4, we show the distribution of stellar mass and SFR 
(M-SFR) for star-forming ETGs. Overall, 
the distributions of our ETG 
and the SDSS SFG samples are similar, and most of our 
star-forming ETGs lie on the `main sequence' (Noeske et al. 
2007; Cano-D\'{i}az et al. 2019). 
The best fit of the MaNGA data at $z\sim 0$ from 
Cano-D\'{i}az et al. (2019) assumed a Salpeter IMF (Salpeter 1955) 
is exhibited by the red dotted-dashed line, and the 
green solid line represents 
the best least-squares fit for our star-forming ETGs. We obtain 
the fit of 
log(SFR)$=(0.74+\pm0.01)$log$(M_{*}/M_{\sun})-(7.22\pm0.08)$, 
and our data have the same slope as the SFGs found in 
Cano-D\'{i}az et al. (2019), showing a slope of 
$0.74\pm0.01$. Also, our fit lies above the 
main sequence from Cano-D\'{i}az et al. (2019), slightly higher 
by $\sim 0.3-0.4$ dex. 
Because the stellar mass/SFR provided by a 
Salpeter IMF is higher $\sim 0.25$ dex than that by a 
Chabrier IMF (Panter et al. 2007; Dutton et al. 2011), the offset 
is mainly due to some IMF selection effect. The galaxy stellar 
mass and SFR are correlated with each other, and 
the close correlation is designated as the main sequence of 
SFGs (Noeske et al. 2007). A significant tendency for 
lower/higher stellar mass ETGs to have lower/higher SFR is shown. 
Compared with the correlation (the Spearman coefficient r=0.60) of 
composite ETGs (Wu 2021), we have the better 
correlation, showing the Spearman coefficient r=0.82. 
This indicates that 
these star-forming ETGs still obey the fundamental 
relation of SFGs, although they actually belong to the ETGs.
In this paper, the stellar mass and SFR are derived by 
using the fiber corrections. If the covering fraction of a galaxy 
is $>0.20$, the systematic and random errors induced by the
aperture effect would be minimized (Kewley, Jansen \& Geller 2005).
The median value of the covering fraction of our sample is 0.38,
and so our result may be rarely influenced by the aperture
effect.

In Table 1, the different parameters of star-forming ETG sample 
are demonstrated. From the MPA-JHU catalogue, we present R.A, Decl, 
stellar masses, redshifts, SFRs, and the fiber and petro flux 
ratio (frac). We also provide the metallicity measurements with 
six abundance indicators for our star-forming ETGs.

\begin{figure*}
\begin{center}
\includegraphics[width=16cm,height=12cm]{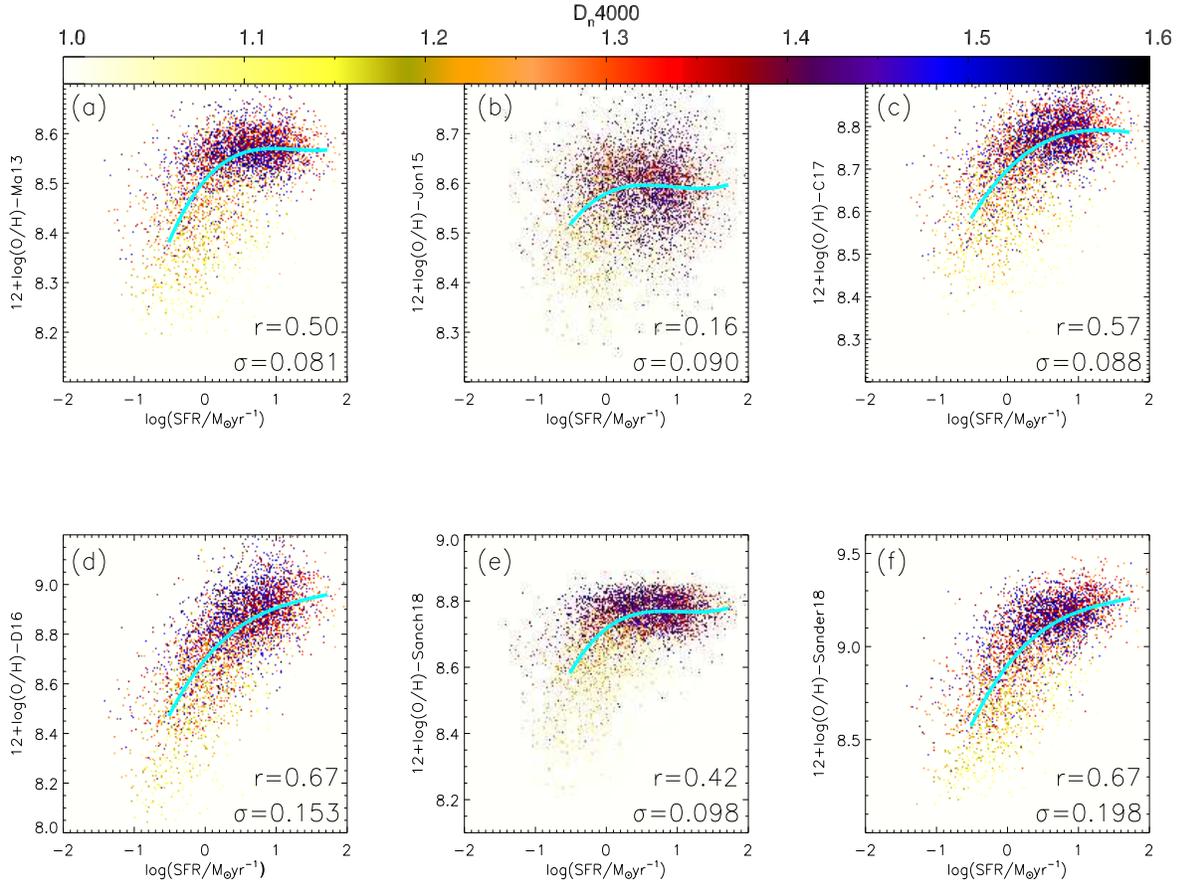}
\caption{Comparison of SFR-Z relations with the colour bar of 
$D_{n}4000$ for star-forming ETGs; these metallicity 
indicators are the same as in Figure 5. The cyan solid lines 
represent a polynomial fit of star-forming ETGs.}
\end{center}
\end{figure*}

\section{Metallicity properties of star-forming early-type galaxies}

Because the star-forming ETGs are selected by using the 
above-mentioned three conditions, we can use various metallicity 
indicators applied for 
SFGs to calculate the metallicity of these ETGs based on 
photoionization models and some empirical methods. In Figure 5, 
we use the six metallicity indicators to present the 
mass-metallicity (MZ) relation for our star-forming ETGs. 
Compared with the stellar mass range of 
$10.0<$log$(M_{*}/M_{\sun})<11.0$ of most composite ETGs in 
Wu (2020), the range of most galaxy stellar masses of our 
star-forming sample significantly increases, from 
log$(M_{*}/M_{\sun})\sim 9.0$ to $\sim 11.3$. 
Compared with the correlation of the MZ relation in composite 
ETGs (Wu 2020), these MZ relations 
of our star-forming ETGs show significantly the positive 
correlation. This may be due mainly to the fact that our ETG 
sample has a wider range of stellar mass.

Figure 5 shows the MZ relations with colour bars 
of EW(H$\alpha$). The colour bar presents the EW(H$\alpha$) values 
per bin in the MZ relations. We can see that star-forming ETGs 
with higher EW(H$\alpha$) values tend to have stronger SF, 
and they often present lower metallicity. 
This is consistent with the 
result of SFGs (e.g., Mannucci et al. 2010). 
In our study, the metallicity is obtained by using 
the SDSS spectra within $3''$ aperture. 
In Iglesias-Paramo et al. (2016), the average difference between 
fiber-based and aperture-corrected metallicities, a typical 
value, is a maximum of 0.047 dex (11\%) for different redshift 
and stellar mass ranges. It depends on the calibrator adopted, 
and it is still within the typical uncertainties of metallicities 
calibrated by empirical estimation, though the averages 
difference is systemically a bias. These results indicate that these 
ETGs almost follow the property of SFGs in the MZ relation, 
and present a good correlation.

\begin{figure*}
\begin{center}
\includegraphics[width=16cm,height=12cm]{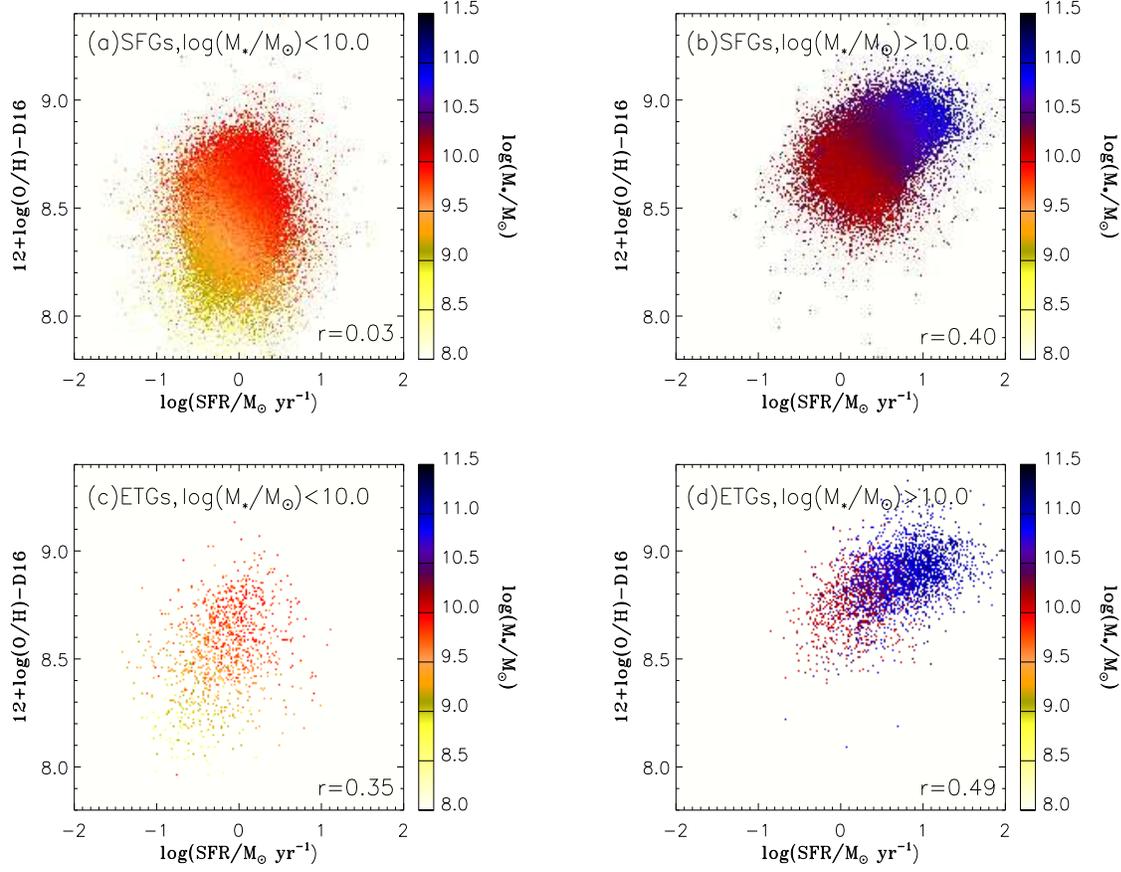}
\caption{Comparison of SFR-Z relations with the 
colour bars of stellar mass for star-forming ETGs and SFGs, 
and these metallicities are calibrated by the D16 
indicator.}
\end{center}
\end{figure*}

In Figure 5, all the green dashed lines correspond to the 
polynomial fits for 148,824 SFGs with $0.023<z<0.3$, obtained 
from the MPA-JHU catalogue for the SDSS DR7 by using the same 
method of sample selection of Wu et al. (2019), except that the 
$\Oii~\lambda \lambda$ 3227, 3229, and $\Nii~\lambda 6584$
have S/N $>2$. Each fit is based on median values of 30 bins, 
and there is 0.1 dex in mass, including more than 200 galaxies. 
The MZ relations for our star-forming ETGs are basically 
consistent with the fit of the above-mentioned large SFG sample.
Figures 5(a)-(f) utilize the metallcity estimators 
of Ma13, Jon15, Curti17, D16, Sanch18, and Sanders18, respectively, 
to calibrate the metallicities of our star-forming 
ETG sample. The MZ relation of Figure 5(a) presents a significant 
correlation, with the Spearman coefficient r=0.57. 
Figures 5(c)-(f) also show the positive correlation, 
and their Spearman coefficients are r=0.61, r=0.70, r=0.55, 
and r=0.67, respectively. In Figure 5, we show each metallicity 
dispersion, calculated in bins of 0.1 dex in $M_*$, for each 
metallicity calibrator. Figure 5(b) does 
not exhibit clearly a positive 
correlation, with the Spearman coefficient r=0.29. 
In addition, these correction coefficients can be used to 
compare the strength of correlations between stellar mass and 
metallicity, but we should consider that the MZ relations of Figure 
5 are sometime not monotonic.

In Wu (2020), because the two metallicity indicators of Sanch18 
and Sander18 show metallicity increasing with N-enrichment, both 
indicators can not calibrate the ETG metallicity. The 
cyan solid lines of Figure 5 show a polynomial fit of star-forming 
ETGs. Each fit is based on median values of 27 bins, 0.1 dex 
in mass, and each bin contains no less than 2 ETGs. We can see that 
most star-forming ETGs locate around the green fit line of 
SFGs in each Figure. Compared with Figures 4(b) and (e) of 
Wu (2020), Figures 5(d)-(f) show that about $67\%$, 
$68\%$, and $66\%$ of star-forming ETGs lie above the median MZ 
relation of large sample SFGs (148,824 SFGs). 
This may be attributed to the effect of $\Nii/Ha$ ratios 
from the so-called diffuse ionized gas (DIG). In a single aperture 
spectra integrating a substantial fraction of a galaxy, such as 
the SDSS data, and in particular for ETGs, it is impossible 
to observe just a single/pure ionisation. In the case of ETGs, 
the effects of shocks and, in particular, diffuse ionisation (from 
the DIG) due to 
hot evolved stars are ubiquitous in this kind of galaxies 
(e.g., Singh et al. 2013; S\'{a}nchez et al. 2014; 
Belfiore et al. 2017). Although the 
EW(Ha) of those contamination - similar to low-ionisation nuclear emission 
line region (LINER), in general- may be lower than that 
of the star formation regions, the $\Nii$/H$\alpha$ ratio 
will increases significantly (Zhang et al. 2017). This means that 
the contamination by this source push the line ratios toward 
some larger values of $\Nii$/H$\alpha$. This way, 
applying an oxygen abundance calibrator that uses the N2 
ratio for an ionised gas with this kind of contamination produces 
an erroneous abundance. This is also true for any calibrator 
using $\Sii$. To provide a more reliable estimation of the 
abundance, it is necessary to decontaminate this component. 
However, this is in general difficult (e.g. Mast et al. 2014; 
Davies et al. 2016; Espinosa-Ponce et al. 2020), and for a 
single fiber spectroscopic data, it is nearly impossible, unless 
certain assumptions are made. This is attributed to the N2S2H$\alpha$, 
N2, and N2O2 indexes from the D16, Sanch18, and Sander18 estimators, 
and we suggest that the three calibrators bring higher metallicity 
measurements. These results indicate that 
star-forming ETGs with low-level SF 
are different from the general population of SFGs, and their 
ionisation may be dominated by SF, but they present a mix of 
other ionisation sources, as they are at least partially 
retired.

In Figure 6, we describe the distributions of SFR and metallicity 
calibrated by the six metallicity indicators. Figure 6(a) uses 
the Ma13 estimator to present 
the relation between SFR and 12+log(O/H)
(SFR-Z), showing significantly a positive correlation, and 
the Spearman coefficient is r=0.50. In SFGs, we use the data 
of Wu et al. (2019) - that is, the metallicity 
calibrated by the O3N2 indicator of Pettini \& Pagel 2004 (PP04), 
and the O3N2 indicator simultaneously utilized by the PP04 and 
Ma13 calibrators - to show the SFR-Z 
relation. We find that these data displays a weak 
correlation, with a Spearman coefficient r=0.34, and are 
consistent with the results found by Yate \& Kauffmann (2014). 
In Figure 6(b), the relation between SFR and 12+log(O/H) is 
shown, and it does not present a correlation, with the Spearman 
coefficient r=0.16. This denotes that the correlation between 
SFR and metallicity may depend on the metallicity indicator.

In Figures 6(c)-(f), the metallicities are calibrated by the 
Curti17, D16, Sanch18, and Sander18,
respectively, and we display their distributions of SFR and 
metallicity. We can see that they present a significant 
correlation, and that their Spearman coefficients are 
r=0.57, r=0.67, r=0.55, r=0.67, respectively. In Figure 6, the 
cyan lines represent a polynomial 
fit of our ETGs. Each fit 
is based on median values of 23 bins, 0.1 dex in log(SFR), 
and each bin includes more than 28 ETGs. 
Figure 6 shows each metallicity dispersion, calculated 
in bins of 0.1 dex in log(SFR), for each metallicity 
calibrator. These results show that 
almost all the relations between SFR and 12+log(O/H) exhibit 
clearly a significant correlation, except the relation 
(calibrated by Jon15 indicator) in Figure 6(b). Compared with 
the weak correlation between SFR and metallicity in composite 
ETG sample (Wu 2021), the two 
parameters of our star-forming ETGs present a significant 
correlation. This indicates that our star-forming ETG sample 
has the better correlation in the SFR-Z relation than the 
SFG and composite ETG samples.

In Figure 6, we also display the SFR-Z relations with
colour bars of $D_{n}4000$. The colour bar presents the
age distribution in the SFR-Z relations. From Figure 6,
star-forming ETGs with higher $D_{n}4000$ index often show higher
metallicity. On this whole, this is consistent with the result
that SFGs having higher $D_{n}4000$ index demonstrate higher
metallicities at a given stellar mass/SFR (Lian et al. 2015;
Wu et al. 2019).

In addition to the differences between the bulk population 
of star-forming ETGs and SFGs, we further explore the differences 
between both types of galaxies. In Figure 7, we use the colour bars 
of stellar mass to show the SFR-Z relations 
of star-forming ETGs and SFGs - the sample of 83,159 SFGs
is obtained by using the method of Wu et al. 2019 and our 
three selection conditions for ensuring 
these SFGs ionised by SF - 
at log$(M_{*}/M_{\sun})<10.0$ and log$(M_{*}/M_{\sun})>10.0$, 
respectively. The metallicity of 
Figure 7 is calibrated by the D16 indicator (the indicator is 
used simultaneously in this study and 
Wu et al. 2019). Figures 7(a) and (b) describe the SFR-Z 
relations for SFGs (35,845 and 47,314 galaxies, respectively) 
at log$(M_{*}/M_{\sun})<10.0$ and 
log$(M_{*}/M_{\sun})>10.0$, respectively, and their Spearman 
coefficients are r=0.03 and r=0.40. In the sample of 83,159 SFGs, 
the SFR-Z relation shows a Spearman coefficient r=0.50. 
In Figures 7(c) and (d), we display the SFR-Z 
distributions for star-forming ETGs (1,188 and 3,427 ETGs, 
respectively) at log$(M_{*}/M_{\sun})<10.0$ 
and log$(M_{*}/M_{\sun})>10.0$, respectively (with their 
Spearman coefficients of r=0.35 and r=0.49). 
In fact, these correlation coefficients that we have utilized 
are valid only when the two variables have a monotonic
relationship. In Figures 7(a) and (b), we 
can see that the metallicity increases with decreasing SFR at a 
fixed stellar mass, log$(M_{*}/M_{\sun} \lesssim 10.5$. 
In Figures 7(c) and (d), we also seem to find the result at a 
given stellar mass. This is consistent with that found in 
Figure 1 of Mannucci et al. (2010), where galaxies with 
log$(M_{*}/M_{\sun}) <10.9$ present a negative correlation. 
From Figures 7(a)-(d), we find that the SFR-Z relation of 
star-forming ETGs have a stronger correlation than that of SFGs.
Although the two types of galaxies have the different sample 
size, in general star-forming ETGs present a better correlation 
for the SFR-Z relation than SFGs.

Wu (2021) found that the 
weak correlation between SFR and 12+log(O/H) may originate from 
metallicity dilution induced by minor mergers. From Figure 7 of 
Wu (2021), we can see that composite ETGs with 
lower stellar mass tend to present lower SFRs. Thus 
they often have a small amount of ISM compared with 
massive ETGs, while minor mergers with gas-rich dwarfs can 
provide the gas, accreted into the center of the galaxy. It is 
easier for the inflow of metal-poor gas induced by minor mergers 
to dilute the metallicity of these ETGs than that of massive 
ETGs (Wu 2021). In SFGs, there are the three primary relations 
involved the stellar mass, SFR, and metallicity. Generally, the 
strongest correlation is the M-SFR relation, and almost 
linear relation between the logarithm of both quantities. Then, 
the tighter correlation is the MZ relation, and finally 
the SFR-Z relation. The SFR-Z relation is, in general, 
considered to be a consequence of the other two relations.

In this study, we show the three relations 
for star-forming ETGs in Figures 4-6, respectively. 
Figure 4 presents the strongest correlation for star-forming ETGs, 
with the Spearman coefficient r=0.82. The stronger correlation 
appears in the MZ relations of Figure 5, compared with the SFR-Z 
relation of Figure 6. We can see that 
the correlations of these primary relations decrease with 
the order of M-SFR, MZ, and SFR-Z. This shows 
that star-forming ETGs also conform to the same sequence of 
the three primary relations for SFGs. In addition to the global 
properties of star-forming ETGs, Figure 7 compares the SFR-Z 
relations with the colour bar of 
stellar mass at log$(M_{*}/M_{\sun})<10.0$ and 
log$(M_{*}/M_{\sun})>10.0$ for the two types of galaxies. 
Star-forming ETGs are basically consistent with the various 
properties of SFGs. Based on the discussion mentioned above, 
we suggest that the correlation of the SFR-Z relation can be 
attributed as a consequence of the other two relations.

The question is whether our results could be affected by the metallicity 
gradient. Although some galaxies show an apparent flattening 
or drop of the metallicity in the central or outer regions, the 
well-known negative abundance gradient has been confirmed by 
different observations (Zaritsky et al. 1994; 
S\'{a}nchez et al. 2012, 2014; Esteban \& 
Garcia-Rojas 2018; S\'{a}nchez 2020). 
In our study, the overestimation of oxygen abundances induced 
by the negative gradient may result in some insignificant changes 
of the corrections and scatters for MZ and SFR-Z relations. This 
should not change the main trends of our results due to the 
entire and small corrections for ETGs.

In addition, from the GALEX results we find that the low-level SF 
in ETGs preferentially occurs at the outskirt of a galaxy 
(Salim et al. 2012; Fang et al. 2012). As SF occurs 
mostly at the ETG outskirts (Gomes et al. 2016), the 
total SFRs of these ETGs may be underestimated. 
Because of low-efficiency SF, our main results cannot be changed by 
the SFRs that are underestimated, although these ETGs may have 
SF at the outskirt. In ETGs, 
the inflow material fueling low-level SF may result in 
their metallicity decreasing (Davis \& Young 2019), while 
the metallicities calibrated for the $3''$ aperture might 
be overestimated. However, these overestimating metallicities do 
not influence our results in statistical significant way.

\section{Summary}

In this work, we have collected observational data of 4,615
star-forming ETGs, derived by cross-matching the 
Galaxy Zoo 1 and SDSS DR7 MPA-JHU emission-line measurements. 
We show the properties of various parameters in star-forming 
ETG sample, and we investigate their metallicities with six 
abundance indicators. Our main results are summarized as 
follows:

1. In our star-forming ETG sample, the distribution of stellar 
mass has a wider range. The median stellar mass of our sample 
is lower by $\sim 0.25$ dex than that of composite ETGs. 
Moreover, the SFR distribution is concentrates at 
$\rm -0.7<log(SFR[M_{\sun}~yr^{-1}])<1.2$, and the median value 
of SFRs of our sample is slightly higher than that shown in 
Davis \& Young (2019).

2. We show a clear trend that lower/higher stellar mass ETGs 
have lower/higher SFRs. We derive the best fit of 
log(SFR)=$(0.74\pm0.01)$log$(M_{*}/M_{\sun})-(7.22\pm0.08)$ 
for our ETG sample, and our data have the same slope as  
that found in Cano-D\'{i}az et al. (2019). This shows that our 
star-forming ETGs locate on the main sequence of SFGs.

3. The MZ relations of star-forming ETGs are shown with six 
abundance indicators. Compared with composite ETGs, most MZ 
relations present significantly a positive correlation in our 
ETGs. We find higher metallicity 
measurements calibrated by the D16, Sanch18, and Sander18 
indicators in star-forming ETGs. 
We suggest that the higher metallicity measurements for
star-forming ETGs can be attributed to larger $\Nii/\rm H \alpha$ ratios
induced by the DIG contamination, although these ETGs are mainly 
ionized by SF.

4. Star-forming ETGs with the low-level SF are similar to 
the general population of SFGs, and their 
ionisation may be dominated by SF, but they present a mix of 
other ionisation sources, as they are at least partially 
retired.

5. The SFR-Z relations are displayed with six abundance 
indicators, showing clearly a positive correlation in five 
indicators. The correlation conforms with the negative 
correlation between the two parameters in SFGs. The correlation 
in ETGs may be a consequence of M-SFR and MZ relations.

\section*{Acknowledgment}
We are very grateful to the anonymous referee for valuable
suggestions and comments, which helped us to improve the paper
significantly. This work is supported by the NSFC (No. 12090041, 
12090040)

\section*{Data Availability}

The data used in this work are publicly available to access 
and download as follows:

(1) the catalogue of the MPA-JHU for the SDSS DR7 is available 
at https://wwwmpa.mpa-garching.mpg.de/SDSS/DR7/.

(2) the NYU-VAGC for the SDSS DR7 is available 
at http://sdss.physics.nyu.edu/vagc-dr7/vagc2/sersic/.

(3) Galaxy Zoo 1 is available 
at https://data.galaxyzoo.org.

\end{document}